\documentclass[11pt, letterpaper, twocolumn]{article}
\usepackage{cite}
\usepackage{graphicx}
\usepackage{lipsum}
\usepackage{multicol}
\usepackage{graphicx}
\usepackage{caption}
\usepackage{refstyle}
\usepackage{amsmath}
\usepackage{amsthm}
\usepackage{amsmath}
\usepackage{amsfonts}
\usepackage{amssymb}
\usepackage{bbm}
\usepackage{amsmath,array}
\usepackage{abstract}
\usepackage{mathtools}
\usepackage{slashbox}
\usepackage[all]{xy}
\usepackage{mathtools}
\usepackage[utf8]{inputenc}
\usepackage[english]{babel}
\usepackage[T3,T1]{fontenc}
\usepackage{fancyhdr}
\usepackage{graphicx} 
\DeclareSymbolFont{tipa}{T3}{cmr}{m}{n}
\DeclareMathAccent{\invbreve}{\mathalpha}{tipa}{16}

\usepackage[left=1in,right=1in,top=1in,bottom=1in]{geometry}
\usepackage{setspace}

\begin{document}

\title{Information Theoretically Secure Encryption Key Generation over Wireless  Networks by Exploiting Packet Errors} 
\author{Amir K. Khandani \\ E\&CE Dept., Univ. of Waterloo, Waterloo, Ontario, Canada; khandani@uwaterloo.ca}

\date{}
 \maketitle   

\noindent
\underline{\bf Abstract:}\footnote{Author acknowledges that the data presented in Figs.~\ref{Fig2},  \ref{Fig4} and \ref{Fig5} are provided by Dr. Ehsan Tavakoli.}
	This article presents a novel method for establishing an information theoretically secure encryption key over wireless channels. It exploits the fact that data transmission over wireless links is accompanied by packet error, while noise terms, and thereby the error events observed by two separate receivers are independent of each other. A number of data packets, with random data, are transmitted from a first legitimate node, say Alice, to a second legitimate node, say Bob. Bob identifies all packets that are received error-free in the first transmission attempt and sends their indices to Alice over a public channel. Then, both Alice and Bob mix the contents of identified packets, e.g., using a hash function, and thereby derive an identical encryption key. Since error events from Alice to Bob is independent of error events from Alice to Eve, the chances that Eve has successfully received all packets used in key generation error-free diminishes as the number of packet increases. In many wireless standards, the first stage in error detection and Automatic Repeat Request (ARQ) is deployed at the PHY/MAC (Physical Layer/Medium Access Control) layer.  In such setups, the first re-transmission is manged by the   PHY/MAC layer without informing higher layers. This makes it impossible  to directly access the information related to packet errors through high-level programming interfaces available to an end-user.  A method is presented for determining packets received error-free in first transmission attempts through high-level programming. Examples are presented in conjunction with an LTE cellular network.


\section{Introduction}
Conventional methods for generating/sharing an encryption key (key encapsulation), excluding Quantum Key Distribution (QKD), rely on three sets of techniques: 
(1) Information theoretical methods that operate based on: (i) adding random noise to data \cite{wire}, or (ii) extracting the information common between two dependent random variables  \cite{common}.   (2) Mathematical methods to construct a one-way function that is hard to invert. (3) Methods motivated by McEliece cryptosystem~\cite{R0} which incorporate randomness in the public key matrix and rely on error correcting codes.  Information theoretical methods, although known to (asymptotically) achieve perfect secrecy, rely on existence results, lacking a clear path to practical realization.  For example, Turbo-like codes are known to approach existence results dealing with channel capacity with a small loss in energy efficiency. This shortcoming  can be handled by an increase in signal power without any additional consequences. However, unlike the case of channel capacity, it is difficult to quantify and/or deal with any remaining gap to relevant existence results in the case of information theoretical security.

Consider a packet wireless data network between legitimate nodes Alice and Bob. To  establish a secure encryption key between the two legitimate nodes, Alice sends a sequence of $N$ User Data Protocol (UDP) packets to Bob. Packets, which contain random data, are equipped with error detection. Upon receiving the $N$ packets, Bob sends the indices of packets received error-free (in the first transmission attempt) to Alice. Then, Alice and Bob use a hash function to mix contents of packets available error-free at both ends. The outcome will be used as an encryption key for secure data exchange, e.g., using Advanced Encryption System (AES). Typically, a key of size 256 bits is sufficient to realize secure transmission using AES. Once the initial key is established, Alice and Bob can send a smaller number of additional packets, called {\em updating packets} hereafter, to refresh the existing key. In each such transmission,  the contents of updating packets received error-free (in the first transmission attempt) are mixed with the current key, e.g.,  using a hash function. To improve key secrecy, Alice and Bob can operate opportunistically and send such updating packets when the channel from Alice to Bob enjoys a high Signal-to-Noise Ratio (SNR). In Section~\ref{Op}, this topic is discussed in relation to a Long Term Evolution (LTE) network.  In addition, 
it is (typically) not feasible for an end-user with limited access to PHY/MAC layer to directly identify which packets are received free of error in the first transmission attempt. A method is presented to overcome this challenge, and examples of test results are presented over an LTE network.

\section{Proposed Method}
Let us assume the packet error probability from Alice to Bob and from Alice to Eve are equal to  $\mathtt{e}_b=1-\mathtt{c}_b$ and $\mathtt{e}_e=1-\mathtt{c}_e$, respectively. In sending $N$ packets from Alice to Bob, the probability that $i$ packets are received correctly is equal to
  \begin{equation}
 {N\choose i} \mathtt{c}_b^{i} (1-\mathtt{c}_b)^{N-i}.
\label{Ave-1}
\end{equation}
The key will be formed from information contents of all $i$ packets received correctly by Bob (in their respective first transmission attempts). As a result, the security will be compromised if  all $i$ packets received correctly by Bob (in a single/first transmission attempt) are also received correctly by Eve. The corresponding probability is
  \begin{equation}
 {N\choose i} \mathtt{c}_{b}^{i}\mathtt{c}_{e}^i (1-\mathtt{c}_b)^{N-i}.
\label{Ave-2}
\end{equation}
It follows that
\begin{eqnarray} \label{Ave-3}
\mathtt{P} & = & \sum_{i=0}^N  {N\choose i} \mathtt{c}_{b}^{i}\mathtt{c}_{e}^i (1-\mathtt{c}_b)^{N-i} \\ \label{Ave-4}
&= & \left(1-\mathtt{c}_b+\mathtt{c}_b\mathtt{c}_e\right)^{N} \\  \label{Ave-5}
&= & \left(1-\mathtt{c}_b\mathtt{e}_e\right)^{N}.
\end{eqnarray}
Security level $\mathsf{SEC}$ is defined as
  \begin{eqnarray} \label{Ave-6p}
\mathsf{SEC} & = & -\log_2\left(\mathtt{P}\right) \\  \label{Ave-6}
&=& -N\log_2\left(1-\mathtt{c}_b+\mathtt{c}_b\mathtt{c}_e\right) \\  \label{Ave7}
&=& -N\log_2 \left(1-\mathtt{c}_b\mathtt{e}_e\right)
\end{eqnarray}
It can be easily established that \ref{Ave7} is a monotonically increasing function 
of  $\mathtt{c}_b$ and a monotonically decreasing 
function of  $\mathtt{c}_e=1-\mathtt{e}_e$. 
This means maximum of $\mathsf{SEC}$ falls on the boundary within  the feasible region
of $\mathtt{c}_b\times \mathtt{c}_e$. In the following, two cases are discussed in more details. 

\subsection{Degraded Channel}
Assuming channel from Alice to Eve is a degraded version of the channel from Alice to Bob, i.e., $\mathtt{c}_b\geq \mathtt{c}_e=1-\mathtt{e}_e$, it follows that the lowest value of $\mathsf{SEC}$ (bottleneck in security level) corresponds to $\mathtt{c}_e=\mathtt{c}_b$, for which, we have
\begin{equation}
\mathsf{SEC}=-N\log_2\left(1-\mathtt{c}_b+\mathtt{c}^2_b\right).
\label{Ave70}
\end{equation}
Expression \ref{Ave70} is maximized $\mathtt{c}_b=1/2$,
resulting in
\begin{equation}
	\hspace{-0.25cm}
\max_{\mathtt{c}_b: \mathtt{c}_b\geq \mathtt{c}_e} \mathsf{SEC}=
\mathsf{SEC}_{\mathtt{c}_b=\mathtt{c}_b=0.5}=
N \log_2 \left(\frac{4}{3}\right).
\label{Ave-8}
\end{equation}
From~\ref{Ave-8}, we conclude 
$\mathsf{SEC}_{\mathtt{c}_b=\mathtt{c}_b=0.5}=256$ is realized for $N=617$ packets. 

\subsection{General Channel} 
\begin{figure}[h]
	\centering
	\hspace{1cm}
	\includegraphics[width=0.51\textwidth]{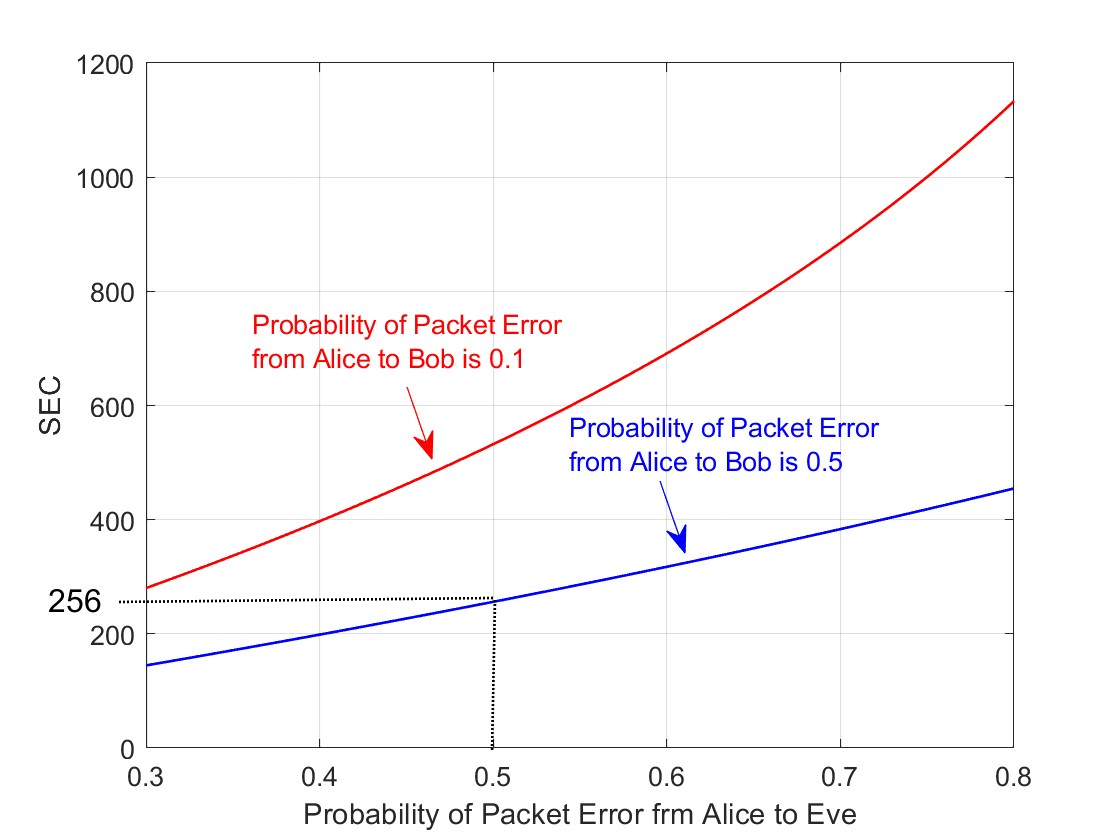}
	\caption{$\mathsf{SEC}$ values for $N=617$ computed using \ref{EQU2}, i.e., $\mathtt{c}_b=0.9$.}
	\label{Fig0}
\end{figure}

Typically, in wireless transmission, system parameters are adjusted such the probability of a packet being received correctly in the first transmission attempt, i.e., prior to any re-transmission due to Automatic repeat request (ARQ), is equal to $\mathtt{c}_b=0.9$. These parameters include power control and/or  adapting of Modulation and Coding Scheme (MCS) index. Replacing in \ref{Ave7}, we obtain

\begin{eqnarray} \label{EQU2} 
	\mathsf{SEC}   &= & -N\log_2 \left(1-0.9\mathtt{e}_e\right)    \\  \nonumber  
	& = & -N\log_2 \left(0.1+0.9\mathtt{c}_e\right).    
\end{eqnarray}
Figure~\ref{Fig0} depicts $\mathsf{SEC}$ values in \ref{EQU2} for $N=617$. 

\section{Example of Usage Scenario} \label{Op}
Method proposed here for establishing encryption keys can be applied to 
a wide range of wireless networks, e.g., cellular, Wireless Local Area Network (WLAN), Bluetooth and various standards deployed in the Internet of Things (IoT). In almost all these standards, it is possible to implement the proposed method through high level programming to send/receive UDP packets. Some deployments also allow for extracting information relevant to opportunistic transmission through high level programming. For example, Figs.~\ref{Fig1} and \ref{Fig2} depict relevant information extracted in a cell phone over a Long Term Evolution (LTE) cellular network.   

A challenge in the application of the proposed method to certain wireless standards, in particular cellular networks, is due to the nature of the underlying Automatic Repeat Request (ARQ) protocol. The problem is that the first stage of ARQ is deployed at the lowest network layer, i.e.,  at the PHY/MAC (Physical Layer/Medium Access Control) layer, where re-transmission for erroneous packets can be neither monitored nor controlled by an end-user. In other words, detecting which packets are received correctly in the first transmission attempt is not feasible, at least not through high level programming.  This is the case even for UDP packets. The problem is that, for packets that are re-transmitted due to ARQ, Eve is also provided with multiple copies, and accordingly, its packet error rate reduces.  In the following, this problem is tackled in the context of a cellular network by separating vast majority of packets that are received correctly in the first transmission attempt, from those that have gone through a low-level re-transmission orchestrated by the PHY/MAC layer. 

To achieve this goal, the sequence of UDP packets used in establishing the key are transmitted at regular time intervals, and their received times (at the layer above PHY/MAC) is recorded through high level programming. Figure~\ref{Fig3} depicts an example of received times for two packets sent consecutively. Let us focus on packets that are received at the higher layer after at most a single PHY/MAC re-transmission. Packets experiencing more than a single PHY/MAC re-transmission are rare and do not affect validity of the following discussions.  From Fig.~\ref{Fig3}, there can be three cases of time gaps (at the receiver side) between successive packets sent at regular time intervals. Case 1 corresponds to Fig.~\ref{Fig3}(a)(b), case 2 corresponds to   Fig.~\ref{Fig3}(c) and case 3 corresponds to  Fig.~\ref{Fig3}(d). Figures~\ref{Fig4} and \ref{Fig5} depict examples of time gaps  measured over an LTE network, where the corresponding mean values are subtracted. Note that the case in Fig.~\ref{Fig4} depicts a connection with a higher variation in packet travel time as compared to that of Fig.~\ref{Fig5}.   It remains to decide which packets are received correctly in the fist transmission attempt.  This is achieved through {\em regularization of received times} explained next.  As mentioned earlier, vast majority of erroneous packets  are received correctly after a single  PHY/MAC layer re-transmission. It will be clear from followup discussions that packets experiencing two or more re-transmission attempts, e.g., Fig.~\ref{Fig3}(b), due to being rare, do not affect the discussions relevant to time regularization presented next. 

\subsection{Regularization of Received Times} \label{TR}

Regularization is achieved relying on the facts that: (i) sequence of UDP packets are transmitted with equal time gaps, and (ii) majority of packets are received correctly in the first transmission attempt, i.e., Fig.~\ref{Fig3}(a). 
Let us rely on notation $\mathsf{t}_i$, $i=1,\ldots,N$ to denote the received time (measured by Bob) for packet  indexed by $i$.   Due to variation in travel times of successive packets $\mathsf{t}_i$, $i=1,\ldots,N$ are not uniformly spaced (are not at equal time intervals). We rely on (minimizing) mean square error criterion to shift (regularize) the values of $\mathsf{t}_i$, $i=1,\ldots,N$ to have equal time gaps. 
Due to a potential difference between clock rates at nodes Alice and Bob, the time gap measured at Bob (using its local oscillator as the reference of time)  is not necessarily equal to the time gap used by Alice. Let us assume the time gap at Bob (if all packets were received in equal time intervals) would be equal to $\mathsf{g}$ (measured based on the clock at Bob). To regularize the packets (order them at regular time intervals), Bob computes the solution  to
\begin{equation}
	\min_{\mathsf{o},\mathsf{g}}\sum_{i=1}^N \left(\mathsf{t}_i-\mathsf{o}-i \mathsf{g}\right)^2
	\label{eq9}
\end{equation}
where $\mathsf{o}$ is an offset and $\mathsf{g}$ denotes the regularized time gap at Bob. Setting the derivatives of \ref{eq9} with respect to $\mathsf{o}$ and $\mathsf{g}$ equal to zero, one can compute $\mathsf{o}$ and $\mathsf{g}$ in closed forms. Adjusting (regularizing) the values of $\mathsf{t}_i$, results in
\begin{equation}
	\mathsf{\hat{t}}_i=\mathsf{o}+i\mathsf{g}.
	\label{eq10}
\end{equation}
Upon regularization, packets that are assigned to regularly spaced points are used in the key generation. 

Note that packets with inter-travel times falling around the center point in Figs.~\ref{Fig4}, \ref{Fig5} are composed of two types: either the two successive packets are both received on time, i.e. Fig.~\ref{Fig3}(a), or are both delayed , i.e., Fig.~\ref{Fig3}(b). For error rates encountered in practice, case in Fig.~\ref{Fig3}(b) has a much lower probability as compared to the case in Fig.~\ref{Fig3}(a). For this reason, one could simply rely on hashing contents of packets falling around the center points in  Figs.~\ref{Fig4},\ref{Fig5} to form the key.
Noting that the final key is affected by contents of all such packets, it is concluded that the small fraction among these where both packets are received with delay (first transmission has been in error for two consecutive packets) will not affect: (i) randomness of the final key, and (ii) independence of the generated key from data available to Eve.  Also, due to hashing, the small number of packets among these that are (potentially) available (error-free) to Eve do not affect the validity of these two conclusions.

\begin{figure}
	\begin{small}
	\begin{center}
		\begin{verbatim}
			Location Information: 
			TAC: Tracking Area Code
			ECI: E-UTRAN Cell Identifier, 28bits, where 
			ECI = CID (16bits) + 65536*RNC (12bits)
			PCI: Physical Cell Identifier
			
			Signal Power:
			RSRP: Reference Signal Strength Power (dBm)
			
			Signal Quality:
			RSRQ: Reference Signal Received Quality (dB)
			RSSNR: Reference Signal SNR (dB)
			CQI: Channel Quality Index, where 
			CQI = 0    --> out of range
			CQI = 1-6  --> 16QAM
			CQI = 7-15 --> 64QAM
		\end{verbatim}
	\end{center}
		\end{small}
	\caption{Example of information that can be extracted through high level programming in a cell phone over an LTE network.}
	\label{Fig1}
\end{figure}

 \begin{figure}[h]
	\centering
	\hspace{1cm}
	\includegraphics[width=0.3\textwidth]{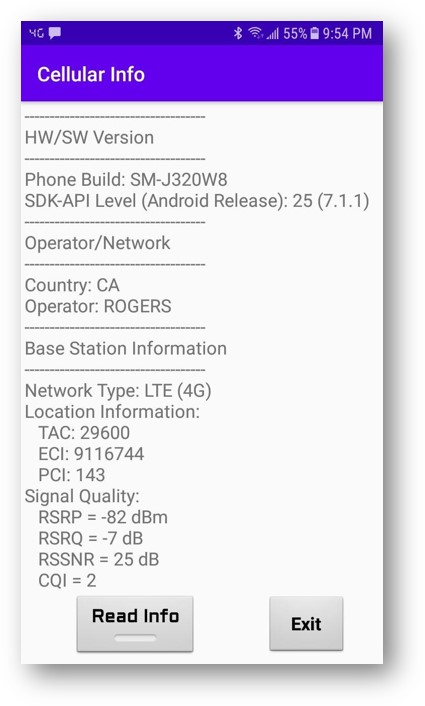}
	\caption{Interface deployed for depicting information extracted in a cell phone operating over an LTE network.}
	\label{Fig2}
\end{figure}

\begin{figure}[h]
	\centering
	\hspace*{-1cm}
	\includegraphics[width=0.4\textwidth]{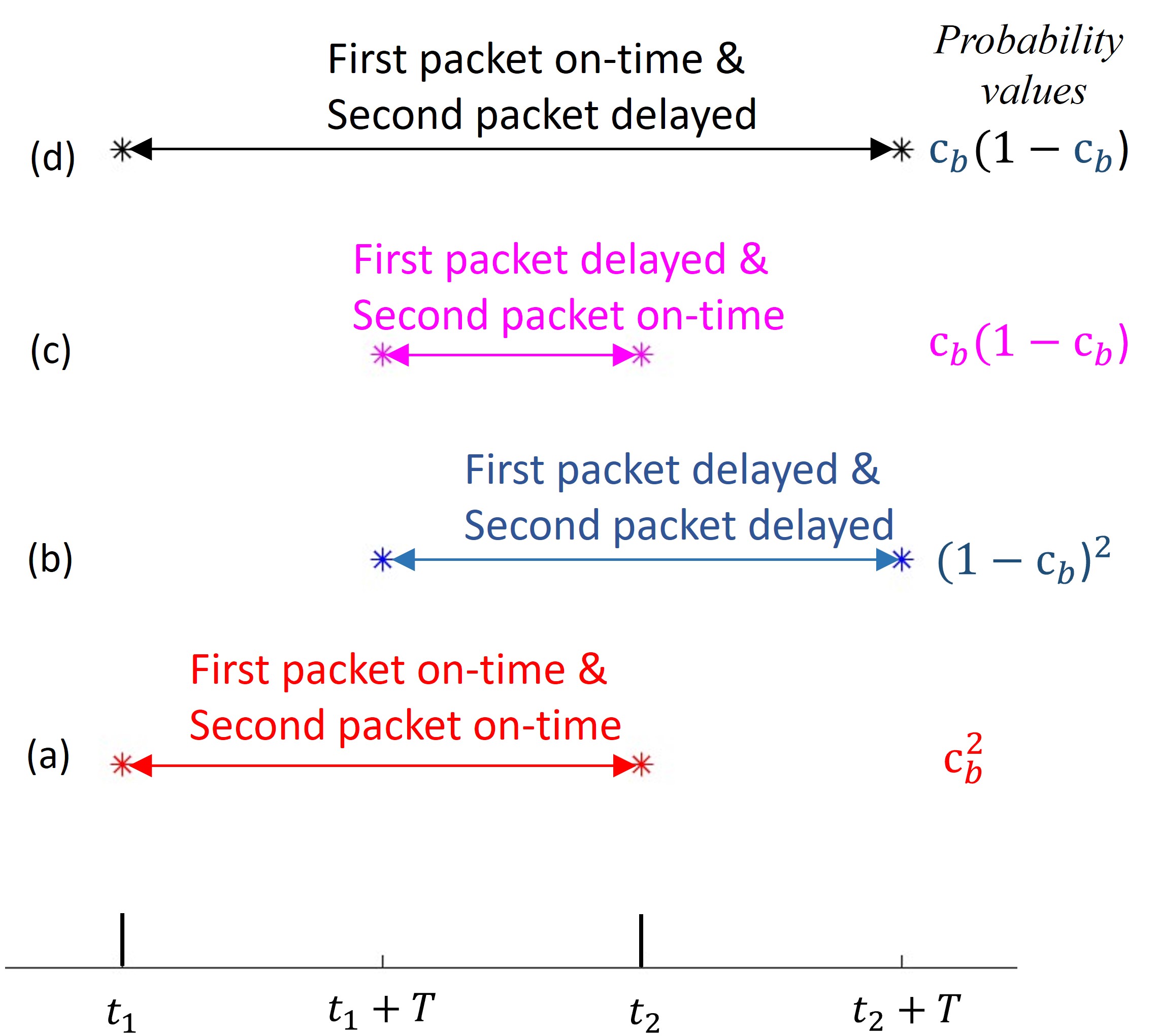}
	\caption{Received times for packets sent at regular time intervals, where $T$ is the round trip time in PHY/MAC ARQ.}
	\label{Fig3}
\end{figure}

 \begin{figure}[h]
	\centering
	\includegraphics[width=0.5\textwidth]{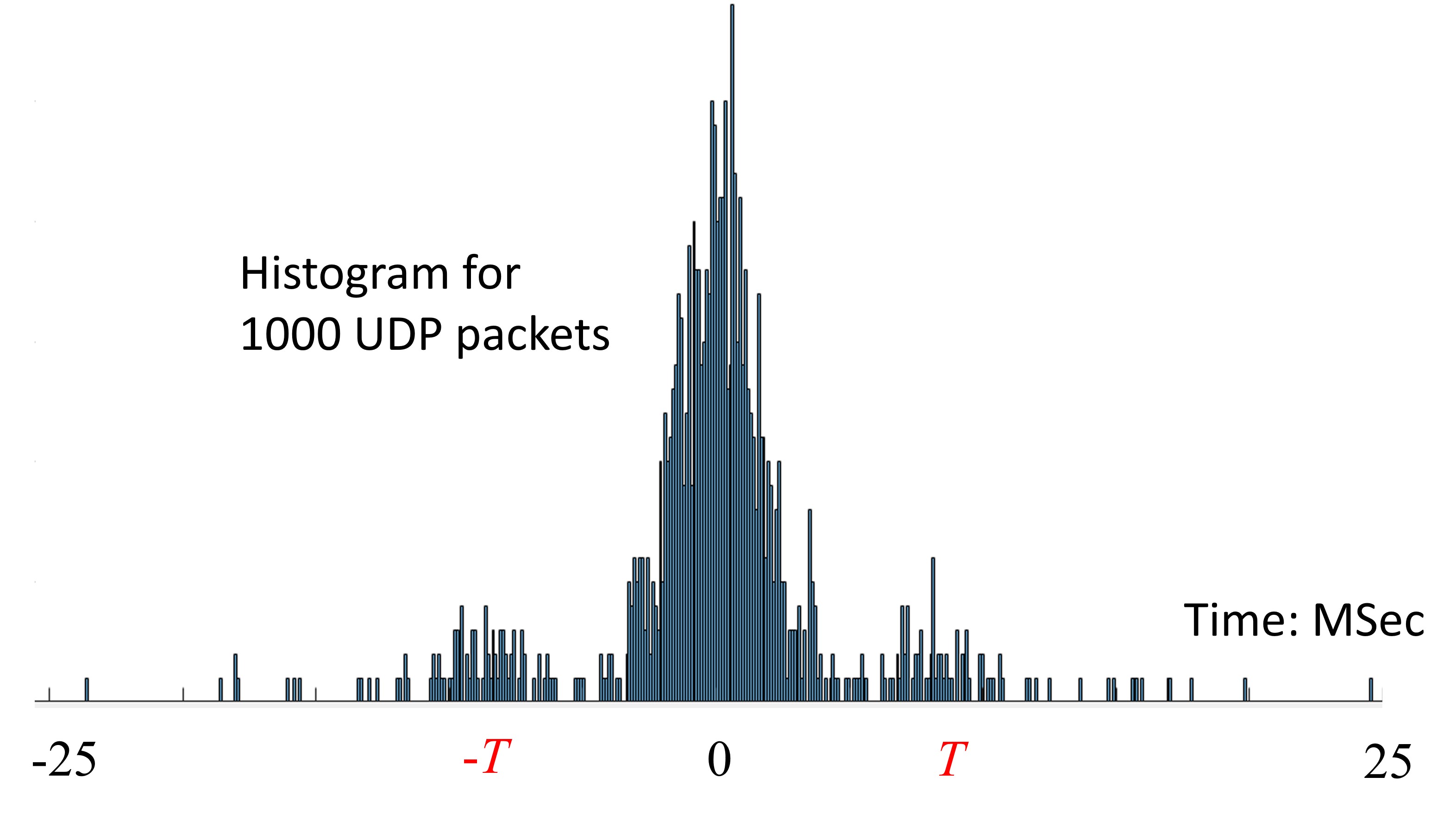}
	\caption{Examples of time gaps  measured over an LTE network, shifted around their respective mean value, where $T$ is the round trip time in PHY/MAC  ARQ.}
	\label{Fig4}
\end{figure}

 \begin{figure}[h]
	\centering
	\includegraphics[width=0.5\textwidth]{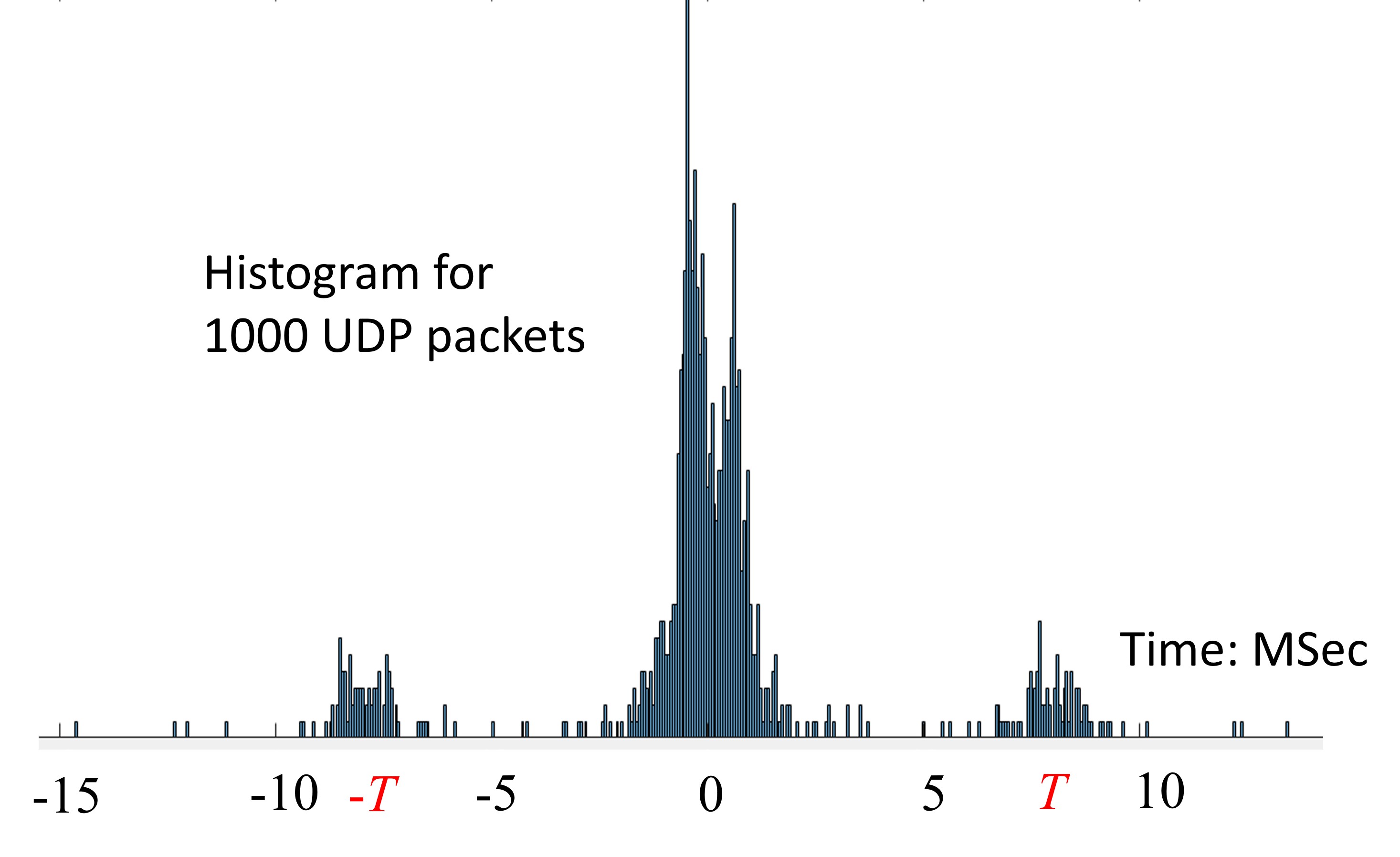}
	\caption{Examples of time gaps  measured over an LTE network, shifted around their respective mean value, where $T$ is the round trip time in PHY/MAC  ARQ.}
	\label{Fig5}
\end{figure}

\begin{small}
	
	\end{small}

\end{document}